\documentclass{moriond}

\usepackage{amsthm,amsfonts,amssymb}
\usepackage{slashed}

\def\be{\begin{equation}}
\def\ee{\end{equation}}
\def\bea{\begin{eqnarray}}
\def\eea{\end{eqnarray}}

\newcommand{\Photo}{}

\begin{document}
\vspace*{4cm}
\title{COHERENT ELASTIC NEUTRINO-NUCLEUS SCATTERING\\
{\footnotesize FIRST CONSTRAINTS / OBSERVATIONS AND FUTURE POTENTIAL}}

\author{Thomas Rink (on behalf of the \textsc{Conus} Collaboration)}

\address{Max-Planck-Institut f\"ur Kernphysik, Saupfercheckweg 1,
69117 Heidelberg, Germany}

\maketitle\abstracts{
The detection of coherent elastic neutrino-nucleus scattering (CE$\nu$NS) opens new possibilities for neutrino physics within and beyond the Standard Model.
Following the initial discovery in 2017, several experimental attempts have emerged allowing this reaction channel to be studied with the full repertoire of modern detection technologies.
As one of several reactor experiments, \textsc{Conus} aims for an observation with antineutrinos emitted from the powerful $3.9$\,GW$_{th}$ reactor of the nuclear power plant in Brokdorf~(Germany).
In particular, the application of ultra-low threshold, high-purity germanium detectors within a sophisticated shield design in close proximity to a nuclear reactor core represents an important step towards high-statistics neutrino detection with small-scale detectors.
In addition to the conventional interaction, typical extensions of the Standard Model neutrino sector can be investigated with data provided from different neutrino sources and several target materials.
Among these, new neutrino interactions as well as electromagnetic neutrino properties are of particular interest. 
This talk gives an overview of existing CE$\nu$NS results and highlights the advantage of using different neutrino sources and target materials.
The example of \textsc{Conus} is used to demonstrate the various capabilities of recent and future CE$\nu$NS measurements.
}


\section{Coherent elastic neutrino-nucleus scattering}

\begin{figure}
	\centering
	\includegraphics[width=0.51\textwidth]{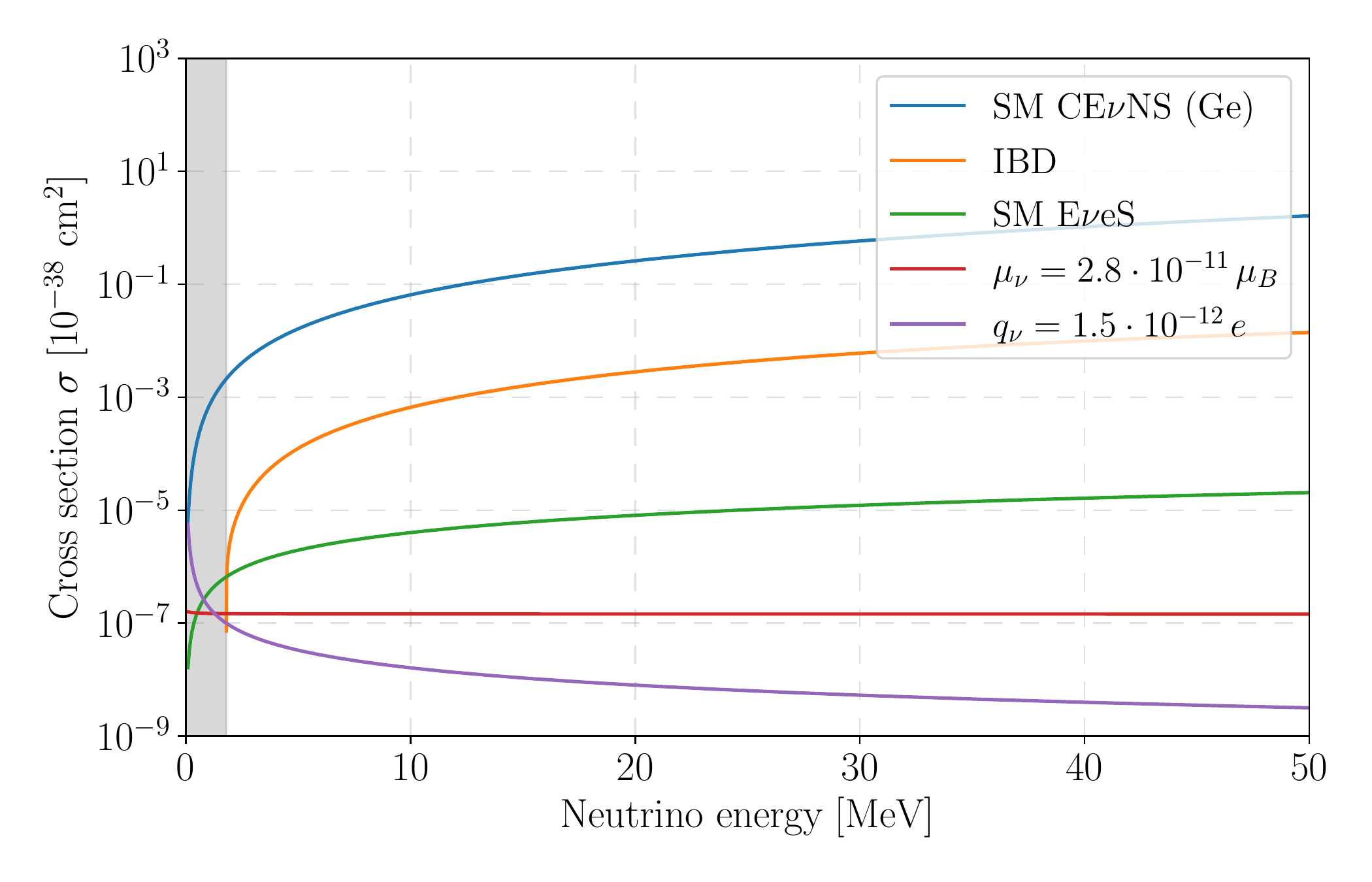}
	\caption[]{Comparison of different neutrino interaction cross section in terms of neutrino energy $E_{\nu}$ for a Germanium target. 
	Coherent elastic neutrino-nucleus scattering (CE$\nu$NS) is about two order of magnitude stronger than inverse beta decay (IBD) and about four orders of magnitude than elastic neutrino-electron scattering (E$\nu$eS).
	Cross sections of electron scattering associated with recent limits on the effective neutrino magnetic moment ($\nu$MM) and a neutrino millicharge ($\nu$MC) are also given for illustration. 
	The figure is taken from Ref.~2.
}
	\label{fig:nu_cc_comp}
\end{figure}

Coherent elastic neutrino-nucleus scattering (CE$\nu$NS) is a flavor-blind and threshold-free interaction channel that is mediated by a weak neutral current.
Shortly after the discovery of the $Z$ boson, it was postulated by Daniel Freedman in 1974~\cite{Freedman:1973yd} and remained undetected for over four decades. 
Within the last five years, first observations were achieved by the \textsc{Coherent} Collaboration at a $\pi$-decay-at-rest ($\pi$DAR) source and experiments located at nuclear reactors already reported first results, i.e.\ \textsc{Connie}, \textsc{Conus} and \textsc{Ncc-1701}.
Further experimental attempts try to measure CE$\nu$NS with different target materials such that it is probed with the full repertoire of modern detection technologies.
The intriguing aspect of CE$\nu$NS is that its coherent nature and the resulting scaling of the cross section with the squared neutron number $N^{2}$, renders it the strongest neutrino interaction within the Standard Model of Particle Physics (SM), if the strict experimental requirements for its detection are met, cf.~Fig.~\ref{fig:nu_cc_comp}.
To guarantee a coherent interaction, the wavelength of the mediating particle must be of the order of the target nucleus, $\lambda_{dB}\sim d_{A}$.
This can be translated into an estimate of the maximally allowed neutrino energy $E_{\nu}\lesssim 80\, A^{-\frac{1}{3}}$\,[MeV], with $A$ being the atomic number of the target nucleus.
Although coherently enhanced, the reaction's observable, the nuclear recoil energy $T_{A}$ of the struck nucleus, scales inversely with the target mass.
Thus, a certain trade-off between the cross section's coherent enhancement ($\sim N^{2}$) and the resulting nuclear recoil energy ($\sim 1/N$) is unavoidable.


The differential CE$\nu$NS cross section is given by
\bea
\frac{d\sigma}{dT_{A}}(T_{A}, E_{\nu})= \frac{G_{F}^{2}}{\pi} \mathcal{Q}^{2}_{W} m_{A} \left(1-
\frac{m_{A} T_{A}}{2 E^{2}_{\nu}}\right) F^{2}(T_{A})\, , \quad \mathrm{where }\ 
	\mathcal{Q}_{W} = \frac{1}{2}\left[(1-4\sin^{2}\theta_{W})Z - N\right]\, ,
\eea
with the Fermi constant $G_{F}$, the nucleus mass $m_{A}$, the nuclear form factor $F^{2}(T_{A})$ and the weak nuclear charge $\mathcal{Q}_{W}$.
The latter is determined by the nucleon composition of the target nucleus, i.e.\ the number of neutrons $N$ and protons $Z$, while the Weinberg angle $\sin^{2}\theta_{W}$ renders it to be mainly determined by the first.
In order to observe CE$\nu$NS, the nuclear recoil energy $T_{A}$ needs to be translated into a ``detectable'' signal, i.e.\ ionization signals in the case of the \textsc{Conus} Germanium (Ge) detectors.
During this conversion (usually referred to as quenching) energy is lost in dissipative processes like heat such that only 10-20\% of the originally stored energy can be read out.
Therefore, quenching puts even stronger constraints on the experimental specifications and energy thresholds below $\sim300$\,eV$_{ee}$ are required for a detection with reactor antineutrinos in a Ge semiconductor detector.


\subsection{Complementarity between different neutrino sources}

In order to satisfy the requirements for a coherently enhanced interaction, neutrinos need to be have energies in the MeV-regime. 
At the moment, the two favored neutrino sources are nuclear reactors (\textsc{Connie}, \textsc{Conus}, \textsc{Ncc-1701}) and $\pi$DAR sources (\textsc{Coherent}, \textsc{Ccm}, \textsc{Ess}), while solar neutrinos are discussed in the context of the recent generation of dark matter direct-detection experiments.
At $\pi$DAR sources, a pulsed proton beam of GeV-energy is shot on a dense target, while subsequent decays of produced pions and muons provide neutrinos of multiple flavors, i.e.\ $\bar{\nu}_{\mu},\nu_{\mu},\nu_{e}$, with energies reaching up to 50\,MeV.
These neutrino energies allow for a large CE$\nu$NS cross section and high nuclear recoil energies, but at the cost of being in a regime where deviations from coherence occur. 
The time correlation between beam dumps and decays allows to suppress background events by a factor of $10^{3}-10^{4}$.
In nuclear reactors, neutrinos are emitted in beta decays of nuclear reaction chains and therefore yield only electron antineutrinos $\bar{\nu}_{e}$. 
With a particle emission of $\sim 10^{20}$\,GW$^{-1}$s$^{-1}$, they are the strongest artificial neutrino sources on Earth and emit antineutrino with energies up to $\sim10$\,MeV.
Although being in a regime where the nuclear form factor can be approximated with unity, the lower neutrino energies consequently lead to lower nuclear recoil energies, which are thus harder to detect.
To take advantage of these huge neutrino fluxes, a position inside the reactor building, close to the core is preferred. 
Usually at these locations, strong safety regulation forbid the use of cryogenic liquids and remote controlling.  
Both neutrino sources have their individual strengths and difficulties, but together they allow us to probe different aspects of CE$\nu$NS at different energies and neutrino flavors.
In this way, they can be viewed as complementary approaches to CE$\nu$NS where information at both sites can be used to gain more knowledge in combined studies.


\subsection{Experimental efforts and results}

For the sake of brevity, we focus here on the four experiments that have already reported first CE$\nu$NS results:

The detectors of the \textsc{Coherent} Collaboration are located at Oak Ridge National Lab (USA) in an experimental corridor close to the facility's interaction point with an overburden of 8\,m of water equivalent (w.e.) and a flux of $4.3\cdot10^{7}$\,cm$^{-2}$s$^{-1}$ at 20\,m distance.
CE$\nu$NS investigations are targeted with four detectors, each having a different energy threshold, mass and target material.
In 2017 and 2019, the collaboration reported successful CE$\nu$NS detection with its CsI[Na] and LAr detectors, where $134\pm22$\,($173\pm48$ predicted)~\cite{COHERENT:2017ipa} and $159\pm43$\,($128\pm17$ predicted)~\cite{COHERENT:2020iec} events at $5\cdot10^{20}$ and $13.7\cdot10^{22}$ protons-on-target have been recorded, respectively.

The reactor experiment \textsc{Connie} is located at the Angra nuclear power plant (Brazil) and is based on silicon CCD technology.
The detector consists of stacked CCDs and exhibits a total mass of 47.6\,g targeting sensitivity to nuclear recoils down to $T_{A}\sim1$\,keV$_{nr}$.
At 30\,m distance to two 3.8\,GW$_{th}$ pressurized water reactors, a neutrino flux of $7.8\cdot10^{12}$\,cm$^{-2}$s$^{-1}$ is achievable.
With data consisting of 2.1\,kg$\cdot$d reactor ON and 1.6\,kg$\cdot$d reactor OFF exposure, the collaboration has limited the event rate related to new physics to be $R_{NP}<40R_{SM}$ at 95\% confidence level (C.L.).\cite{CONNIE:2019swq}
Further, the \textsc{Ncc-1701} experiment is located at 10.3\,m distance to the 2.96\,GW$_{th}$ boiling water reactor of the Dresden nuclear power plant (USA).
With a 3\,kg p-type point-contact Ge detector, CE$\nu$NS is probed down to the recoil energy region of 1-2\,keV$_{nr}$, while at site a neutrino flux of $5\cdot10^{13}$\,cm$^{-2}$s$^{-1}$ is expected.
In the beginning of 2022, ``suggestive evidence'' for a signal detection has been claimed by the collaboration with 289\,kg$\cdot$d reactor ON and 75\,kg$\cdot$d reactor OFF exposure.\cite{Colaresi:2022obx}

The \textsc{Conus} experiment is located at 17.1\,m distance to the pressurized water reactor of the Brokdorf nuclear power plant (Germany). 
The experimental site exhibits an overburden of 10-45\,m w.e. (depending on the azimuthal angle) and a neutrino flux of $2.3\cdot10^{12}$\,cm$^{-2}$s$^{-1}$ is expected.
The shield consists of active and passive components that allow background reduction factors of $10^{3}-10^{4}$ such that a background level of $\mathcal{O}(10)$\,keV$^{-1}$d$^{-1}$kg$^{-1}$ in the region of interest is achieved.
Further, critical reactor-correlated background contributions are proven to be under control.
The \textsc{Conus} set-up applies four kg-size p-type point-contact detectors with (ionization) energy thresholds of $\sim300$\,eV$_{ee}$ (corresponding to $T_{A}\sim2$keV$_{nr}$).
Emphasis was put on a special low-background design and electric cryocooling is chosen due to safety restriction inside the reactor building.
In order to fulfill the strict laboratory conditions close to the reactor core, site monitoring has been steadily improved.
In 2020, the \textsc{Conus} Collaboration constrained the CE$\nu$NS signal rate to be $R_{SM}<0.4$\,kg$^{-1}$d$^{-1}$ at 90\% C.L. (for the currently favored quenching description: (modified) Lindhard model $k=0.16$~\cite{Scholz:2016qos,Bonhomme:2022lcz}) with a data set consisting of 249\,kg$\cdot$d reactor ON and 59\,kg$\cdot$d reactor OFF exposure.\cite{CONUS:2020skt}


\section{BSM investigations of existing data sets}

In the following, we will give an overview of BSM models than can be tested with data collected from $\pi$DAR or reactor experiments. 
The presented results of the \textsc{Conus} Collaboration rely on data sets with exposures of 209\,kg$\cdot$d reactor ON and 38\,kg$\cdot$d reactor OFF for CE$\nu$NS and 649\,kg$\cdot$d reactor ON and 94\,kg$\cdot$d reactor OFF for E$\nu$eS. 
The $\nu$MM investigation via E$\nu$eS uses an even larger data sets with 689\,kg$\cdot$d reactor ON and 131\,kg$\cdot$d reactor OFF exposure.


\subsection{Non-standard interactions}


\begin{figure}
	\centering
	\includegraphics[width=0.46\textwidth]{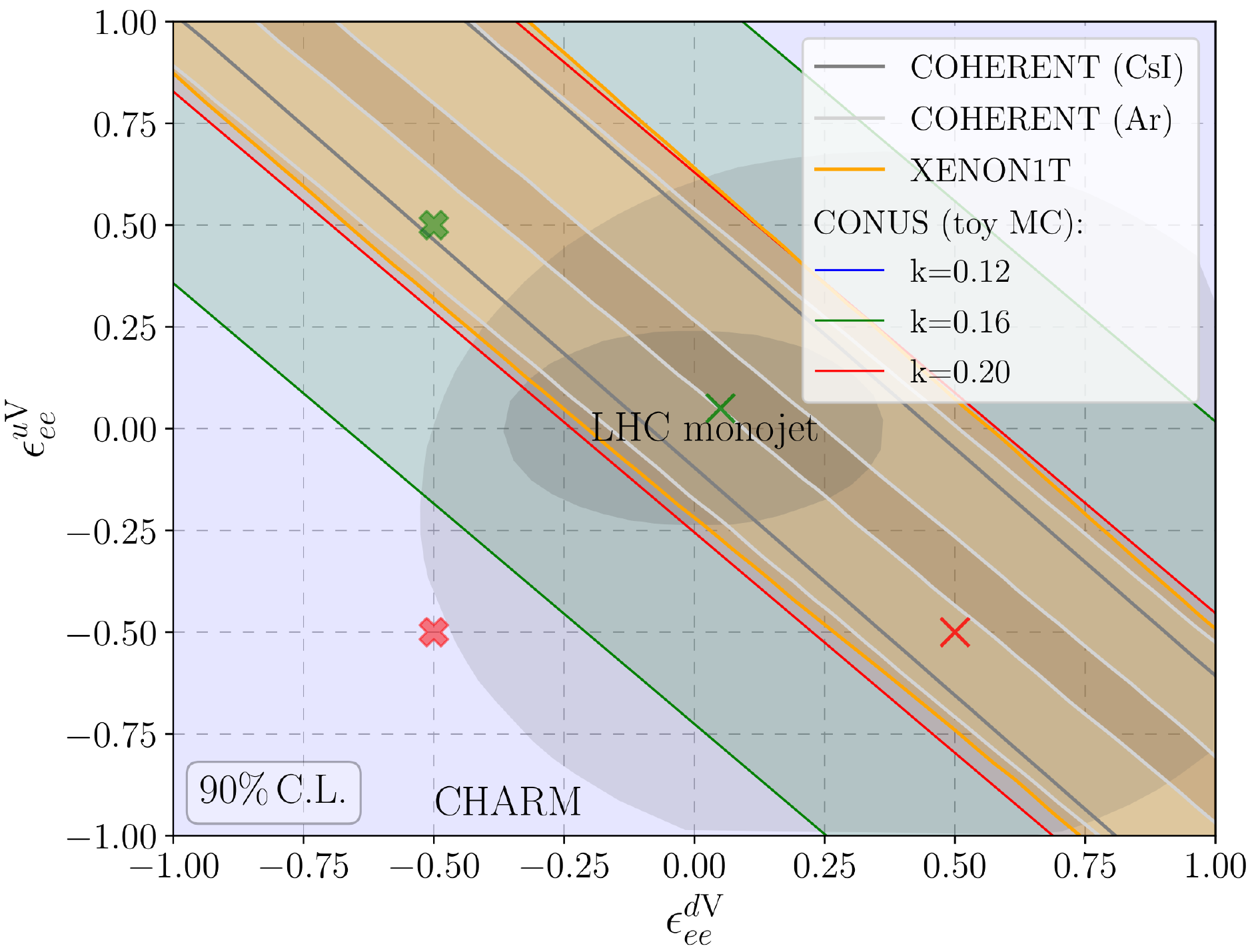}
	\hfill
	\includegraphics[width=0.46\textwidth]{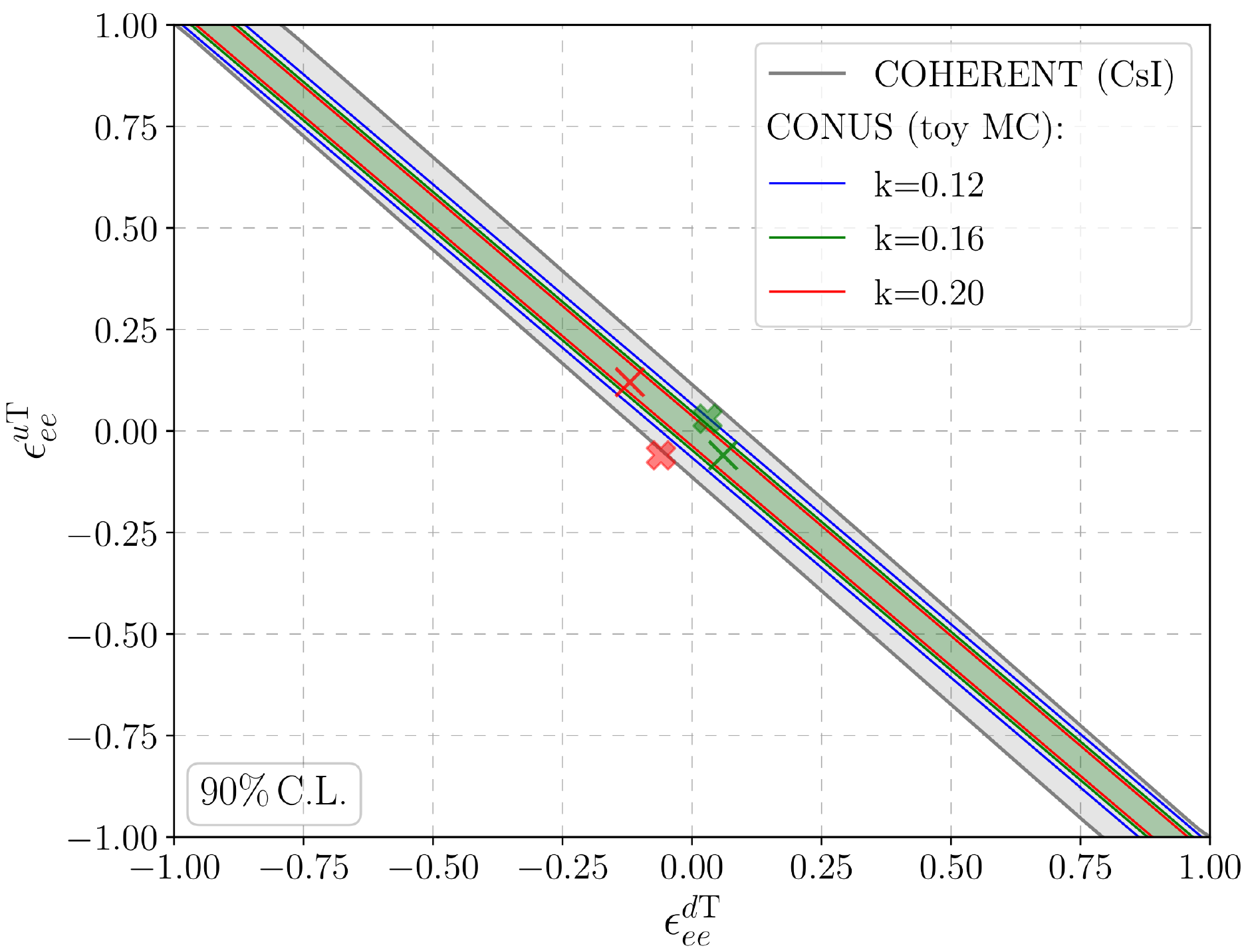}
	\caption[]{Limits on up and down quark couplings of non-standard interactions for a vector-type (left) and tensor-type operator (right).
		Illustrated are the allowed regions for various CE$\nu$NS and non-CE$\nu$NS experiments, while the limits of \textsc{Conus} are given for different quenching parameters ($k$ parameters of the modified Lindhard model). 
		The currently favored quenching behavior is given by $k=0.16$ (green). 
		The figures are taken from Ref.~11.
	}
	\label{fig:nsis}
\end{figure}


At first, we focus on a rather model-independent test of BSM physics.
In the framework of non-standard interactions (NSIs), new physics contributions are expected to arise from the UV scale such that they manifest themselves at low-energy in terms of effective operators.\cite{Lindner:2016wff}
Here, we test two of such operators: a vector-type and a tensor-type operator.

For the investigation of vector-type NSIs, we assume the existence of an operator that exhibits the same chiral structure as the SM interaction, i.e.
\bea\label{eq:operator_vector_nsi}
\mathcal{O}^{q \mathrm{V}}_{\mathrm{NSI}}=\left(\bar{\nu}_{\alpha} \gamma^{\mu} P_{L} \nu_{\beta}
\right)\left(\bar{q} \gamma_{\mu}Pq  \right) + h.c. \, ,
\eea
with the first generation of quarks $q=\{u,d\}$, the chiral projection operators $P=\{P_{L}, P_{R}\}$ and the flavor indices $\alpha,\beta$.
The corresponding couplings $\epsilon$ can be absorbed into the original weak nuclear charge,
\bea\label{eq:nuclear_charge_vector_nsi}
\mathcal{Q}_{\mathrm{NSI}}^{\mathrm{V}} =& \left( g_{V}^{p} + 2 \epsilon^{u\mathrm{V}}_{\alpha \alpha} 
+ \epsilon^{d\mathrm{V}}_{\alpha \alpha} \right) Z + \left(g_{V}^{n} + \epsilon^{u\mathrm{V}}_{\alpha \alpha} + 2 \epsilon^{d\mathrm{V}}_{\alpha \alpha}\right) N \nonumber \\ 
&+ \sum_{\alpha, \beta} \left[ \left( 2\epsilon^{u\mathrm{V}}_{\alpha\beta} + \epsilon^{d\mathrm{V}}_{\alpha\beta} \right)Z + \left( \epsilon^{u\mathrm{V}}_{\alpha \beta } + 2 \epsilon^{d\mathrm{V}}_{\alpha \beta} \right)N  \right]\, ,
\eea
such that different targets/isotopes are needed in order to break occurring degeneracies.
Due to the same final states, this operator can interfere with the SM CE$\nu$NS interaction and may even lead to a reduced number of signal counts.
Bounds an vector NSIs of electron-type from different neutrino sources and target materials are depicted in Fig.~\ref{fig:nsis}.
Since the strength of the new couplings is usually given in terms of the Fermi constant $G_{F}$, one can convert the obtained bounds on the coupling $\epsilon$ into a scale below which new physics can be excluded.
Taking the maximal deviation from zero along the main diagonal, which can be motivated by the assumption of universal couplings to the first quark generation, one obtains the scales $\Lambda_{\mathrm{NP}}>100$GeV (\textsc{Conus}, $k=0.16$) and $\Lambda_{\mathrm{NP}}>240$GeV (\textsc{Coherent}, LAr), reflecting that \textsc{Conus} data are currently not competitive to current $\pi$DAR data.

Further, we test for more exotic BSM physics in terms of a tensor-type operator of the form 
\bea\label{eq:operator_tensor_nsi}
\mathcal{O}^{q \mathrm{T}}_{\mathrm{NSI}}=\left(\bar{\nu}_{\alpha} \sigma^{\mu \nu} \nu_{\beta} \right)\left(\bar{q} \sigma_{\mu \nu} q  \right) + h.c. \, ,
\eea
with $\sigma^{\mu \nu}=\frac{i}{2}[\gamma^{\mu}, \gamma^{\nu}]$ and the first generation of quarks $q=\{u,d\}$.
The corresponding cross section is given by \footnote{While the investigated coherently enhanced tensor contribution appears at subleading order in the non-relativistic expansion, there is also a non-coherent contribution at leading order that leads to a new nuclear form factor. A detailed study of this tensor operator in the multipole decomposition and other BSM contributions can be found in Ref.~12.}
\bea\label{eq:cross_section_tensor_nsi_cenns}
\left(\frac{d\sigma}{dT_{A}}\right)_{\mathrm{tNSI}} = \frac{4 m_A G^{2}_{F}}{\pi} \left[\left( 2\epsilon^{u \mathrm{T}}_{\alpha\beta} + \epsilon^{d \mathrm{T}}_{\alpha\beta} \right) Z +  \left( \epsilon^{u \mathrm{T}}_{\alpha\beta} + 2\epsilon^{d \mathrm{T}}_{\alpha\beta} \right) N\right]^{2} \, \left(1- \frac{m_{A} T_{A}}{4 E^{2}_{\nu}}\right)\, ,
\eea
again with the mass, neutron and proton number of the target nucleus $m_{A}$, $N$ and $Z$, respectively. 
Since the chiral structure of this interaction differs from SM one, there is no interference and this BSM channel only contributes additional events.
Moreover, this interaction exhibits a higher kinematic cut-off, which allows it to extend to higher energies.
This renders quenching less relevant such that the obtained bounds are determined primarily by the detectors' energy threshold and the achieved background levels.
The obtained limits on the couplings to up and down quarks can be found in Fig.~\ref{fig:nsis}.
Here, \textsc{Conus} is able to set competitive bounds.
However, the acquired data sets are not able to break the degeneracy among both couplings, yet.
By taking again the maximal deviation from zero coupling along the diagonal, we can exclude new physics contributions of the introduced tensor-type operator below  $\Lambda_{\mathrm{NP}}\gtrsim360$\,GeV (CONUS, $k=0.16$).

In conclusion, we highlight that sub-percentage precision on the $\epsilon$ parameters translates to a sensitivity on new physics located at the TeV-scale.
Thus, small-scale CE$\nu$NS experiments might represent a complementary probe of this scale that is usually tackled by \textsc{Lhc} experiments.\cite{Lindner:2016wff}


\subsection{Light mediator searches}

One can further search for new mediating particles that lead to spectral distortions in the recorded recoil spectrum at lowest energies.
To investigate these, we apply the more general framework of simplified models, which allows us to lay aside details of model building, but include important kinematic features.\cite{Cerdeno:2016sfi}

At first, we assume the existence of a CP-even, massive and real scalar field $\phi$ and a right-handed (sterile) neutrino $\nu_{R}$ with a Yukawa term that couples left-handed and right-handed neutrino components
The corresponding Lagrangian induces neutrino interactions with the first generation of quarks $q=\{u,d\}$ and electrons and is given by the following expression
\bea\label{eq:lagrangian_light_scalar}
\mathcal{L}_{\phi} = \phi\left( g^{q \mathrm{S}}_{\phi} \bar{q}q + g^{e \mathrm{S}}_{\phi} \bar{e}e + g^{\nu \mathrm{S}}_{\phi} \bar{\nu}_{R} \nu_{L}  + h.c. \right) 
- \frac{1}{2} m^{2}_{\phi} \phi^2\, ,
\eea
with the scalar couplings $g_{\phi}^{x \mathrm{S}}$ to the SM fermions $x=\{q,e,\nu \}$ and the scalar mass $m_{\phi}$.
Since both scalar interactions, CE$\nu$NS and E$\nu$eS respectively, flip the neutrino's chirality, they do not interfere with the SM channels and are purely additive contributions, whose cross sections are given by 	
\bea
\left(\frac{d\sigma}{dT_{A}}\right)_{\phi} = \frac{(g^{\nu \mathrm{S}}_{\phi} \mathcal{Q}_{\phi})^{2} m_{A}^{2} T_{A} }{4\pi E_{\nu}^{2} (2 m_{A} T_{A}  + m_{\phi}^{2})^{2}}\, , \quad \quad \quad \left(\frac{d\sigma}{dT_{e}}\right)_{\phi} = \frac{ (g_{\phi}^{\nu \mathrm{S}} g^{e \mathrm{S}}_{\phi})^{2}\,  m_{e}^{2} T_{e} }{4\pi E_{\nu}^{2} (2 m_{e} T_{e}  + m_{\phi}^{2})^{2}}\, ,
\eea
with the scalar equivalent of the weak nuclear charge $\mathcal{Q}_{\phi}$ and the nuclear and electron masses and recoil energies, $m_{A,e}$ and $T_{A,e}$, respectively.
To further reduce the parameter space, we assume in the following universal couplings to all SM fermions such that the model's parameter space is spanned by only two parameters, i.e.\ the scalar mass $m_{\phi}$ and the coupling $g_{\phi}$.
With this assumption, the scalar charge of the nucleus simplifies to $\mathcal{Q}_{\phi} \rightarrow g_{\phi} (14N+15.1Z)$.
	

\begin{figure}
	\centering
	\includegraphics[width=0.46\textwidth]{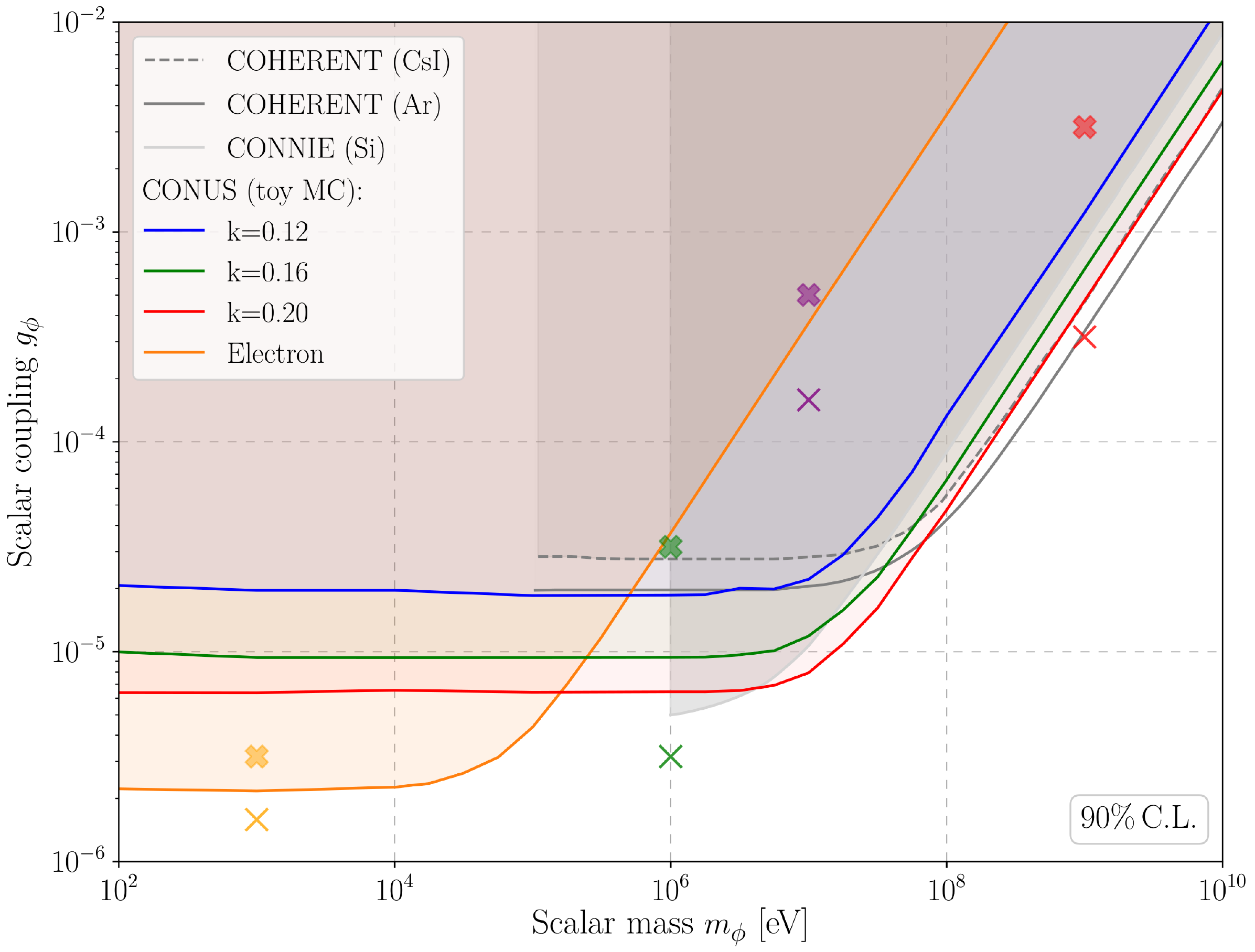}
	\hfill
	\includegraphics[width=0.46\textwidth]{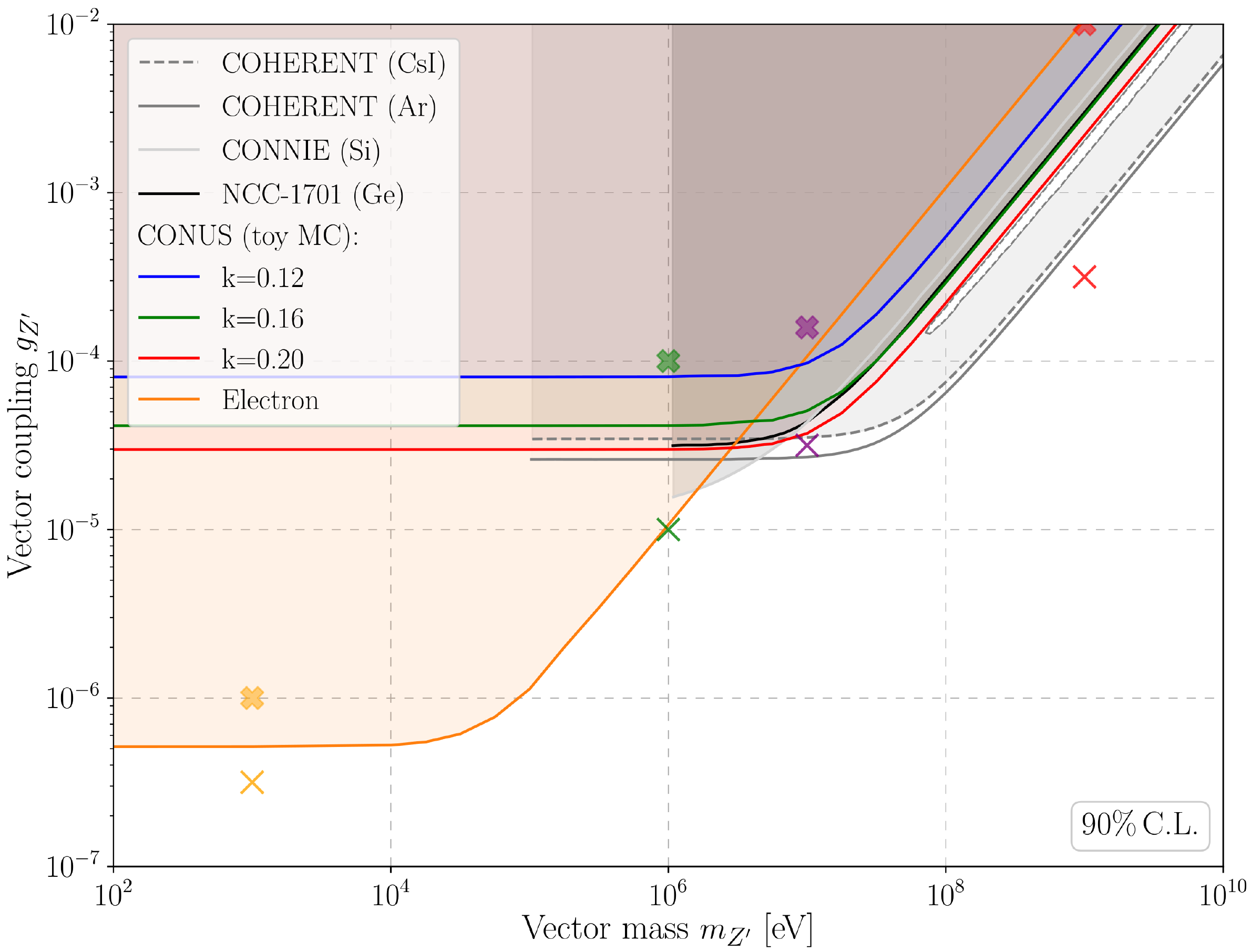}
	\caption[]{Limits on the parameter space of a light scalar (left) and a light vector mediator (right).
		The results of various CE$\nu$NS experiments are presented, while CE$\nu$NS limits of \textsc{Conus} are given for different quenching parameters ($k$ parameters of the modified Lindhard model). 
		The currently favored quenching behavior is given by $k=0.16$ (green). 
		Limits from E$\nu$eS are deduced from a larger data set and indicated in orange.
		The figures are taken from Ref.~11.
	}
	\label{fig:mediators}
\end{figure}


We also test for the existence of a light vector boson that mediates a new force related to an additional $U(1)$ gauge group. 
In this case, we do not include any additional particles, but assume (at least) the first generation of quarks $q=\{u,d\}$ and the electron to be charged under the new symmetry and a Lagrangian to be of the form 
\bea\label{eq:lagrangian_light_vector}
\mathcal{L}_{Z'} = Z'_{\mu} \left( g^{\nu \mathrm{V}}_{Z'} \bar{\nu}_{L} \gamma^{\mu} \nu_{L}
+  g^{e \mathrm{V}}_{Z'} \bar{e} \gamma^{\mu} e  + g^{q \mathrm{V}}_{Z'} \bar{q} \gamma^{\mu} q \right) 
+ \frac{1}{2} m^{2}_{Z'} Z'_{\mu}Z'^{\mu}\, ,
\eea	
with the vector couplings $g^{x \mathrm{V}}_{Z'}$ to the SM fermions $x=\{q,e,\nu \}$ and the boson mass $m_{Z'}$.
Similar to the case of vector NSIs, the new interaction has the same chiral structure as the SM channel and, thus, interference is possible.
The corresponding cross sections for CE$\nu$NS and E$\nu$eS (including the SM contribution) exhibit the form
\bea\label{eq:cross_section_light_vector_cenns}
\left(\frac{d\sigma}{dT_{A}}\right)_{\mathrm{CE\nu NS\,+\,Z'}} = \left( 1	+ \frac{g^{\nu \mathrm{V}}_{Z'}}{\sqrt{2}G_{F}} \frac{\mathcal{Q}_{Z'}}{\mathcal{Q}_{W}} \frac{1}{2 m_{A} T_{A} + m^{2}_{Z'}} \right)^{2}\left(\frac{d\sigma}{dT_{A}}\right)_{\mathrm{CE\nu NS}}\, , 
\eea
\bea\label{eq:cross_section_ligth_vector_nu_e}
\left(\frac{d\sigma}{dT_{e}}\right)_{\mathrm{E\nu eS\,+\,Z'} }=\left(\frac{d\sigma}{dT_{e}}\right)_{\mathrm{E\nu eS}} 
+ \frac{\sqrt{2} G_{F} m_{e} g_{V} g_{Z'}^{\nu \mathrm{V}} g_{Z'}^{e \mathrm{V}} }{\pi (2m_{e}T_{e}  + m_{Z'}^{2})} 
+ \frac{m_{e} (g_{Z'}^{\nu \mathrm{V}} g_{Z'}^{e \mathrm{V}})^{2} }{2\pi (2m_{e}T_{e}  + m_{Z'}^{2})^{2}}\, ,
\eea
with the vector equivalent of the weak nuclear charge $\mathcal{Q}_{Z'}$.
As before, we assume universal coupling to the SM fermions such that the vector charge simplifies to $\mathcal{Q}_{Z'} \rightarrow 3\left( Z+N \right)g_{Z'}$ and the parameter space is reduced to the new mediator's mass $m_{Z'}$ and one universal coupling $g_{Z'}$.

The obtained limits for both mediator investigations and interaction channels are shown in Fig.~\ref{fig:mediators} in comparison to other CE$\nu$NS experiments. 	
There are two characteristic regions for both interaction channels: a flat region, where $2m_{A,e}T_{A,e}>>m_{\phi,Z'}^{2}$ holds, and a slope with $2m_{A,e}T_{A,e}<<m_{\phi,Z'}^{2}$.
Here, the complementarity between CE$\nu$NS experiments using reactor antineutrinos and neutrinos from $\pi$DAR sources becomes evident: due to their characteristic neutrino energy they are sensitive to different mediator mass regions.
Hence, reactors probe lower masses while $\pi$DAR experiments are most sensitive to higher ones.
Further, the huge antineutrino flux at reactor site allows to deduce even stronger bounds via E$\nu$eS for small mediator masses.


\subsection{Electromagnetic properties of the neutrino}

Finally, we illustrate what information can be drawn from CE$\nu$NS data sets in the context of electromagnetic neutrino properties.\cite{Giunti:2014ixa}
In the SM, the neutrino is a left-handed, neutral fermion.
Thus, electromagnetic properties can only arise via loop processes which could allow BSM physics to enter as well.
Consequently, related quantities, e.g.\ a finite $\nu$MM or $\nu$MC, can be used to probe new physics.
Since properties of such quantities depend also on the neutrino's fermion nature, the question whether neutrinos are Dirac or Majorana fermions might be addressed.  
Due to the higher flux at reactor sites, the effective electron $\nu$MM can be investigated via both CE$\nu$NS and E$\nu$eS, while the multi-flavor neutrino content of $\pi$DAR sources allows to probe both electron and muon as well as transition $\nu$MMs via CE$\nu$NS.
The cross sections of the corresponding $\nu$MM-induced interactions are given by
\bea\label{eq:cross_section_nmm}
\left(\frac{d\sigma}{dT_{e}}\right) = C \left(\frac{1}{T_{e}}- \frac{1}{E_{\nu}}\right)\left(\frac{\mu_{\nu_{e}}}{\mu_{B}}\right)^{2}\, , \quad \quad 
\left(\frac{d\sigma}{dT_{A}}\right) =C Z^{2}\left(\frac{1}{T_{A}}- \frac{1}{E_{\nu}}\right)\left(\frac{\mu_{\nu_{e,\mu}}}{\mu_{B}}\right)^{2}F^{2}(T_{A}) \, ,
\eea
with $C=(\pi \alpha^{2}_{\mathrm{em}})/m^{2}_{e}$, the proton number of the target nucleus $Z$ and the nuclear form factor $F^{2}(T_{A})$.
Note that both cross sections scale inversely with the recoil energy $T_{e,A}$, giving an advantage to detectors with very low energy thresholds.
The \textsc{Conus} data sets constrain the effective electron $\nu$MM via E$\nu$eS to $\mu_{\nu_{e}} < 7.5 \cdot 10^{-11}\, \mu_{B}$ (90\%~C.\,L.), which lies in the range of typical reactor bounds.
Assuming a null result, the different scaling in the cross sections of $\nu$MM and $\nu$MC allows to convert this limit into a corresponding limit for a neutrino millicharge. 
In doing so, the \textsc{Conus} data yield a limit of $|q_{\nu_{e}}| < 3.3 \cdot 10^{-12}\ e$ (90\%~C.\,L.).\cite{CONUS:2022qbb}
With \textsc{Coherent} data sets $\nu$MMs have been investigated with the CE$\nu$NS channel. 
Although coherently enhanced (reflected by the squared proton number), the obtained bounds are generally one to two orders of magnitude weaker than current limits from a reactor site, $\mu\sim\mathcal{O}(10^{-10}-10^{-9})\mu_{B}$.
Dedicated investigations of $\nu$MCs at $\pi$DAR sources yield bounds of $|q_{\nu}| \lesssim \mathcal{O}(10^{-8})e$.\cite{Cadeddu:2020lky}


\section{Opportunities for CE$\nu$NS experiments}


\begin{figure}
	\centering
	\includegraphics[width=0.51\textwidth]{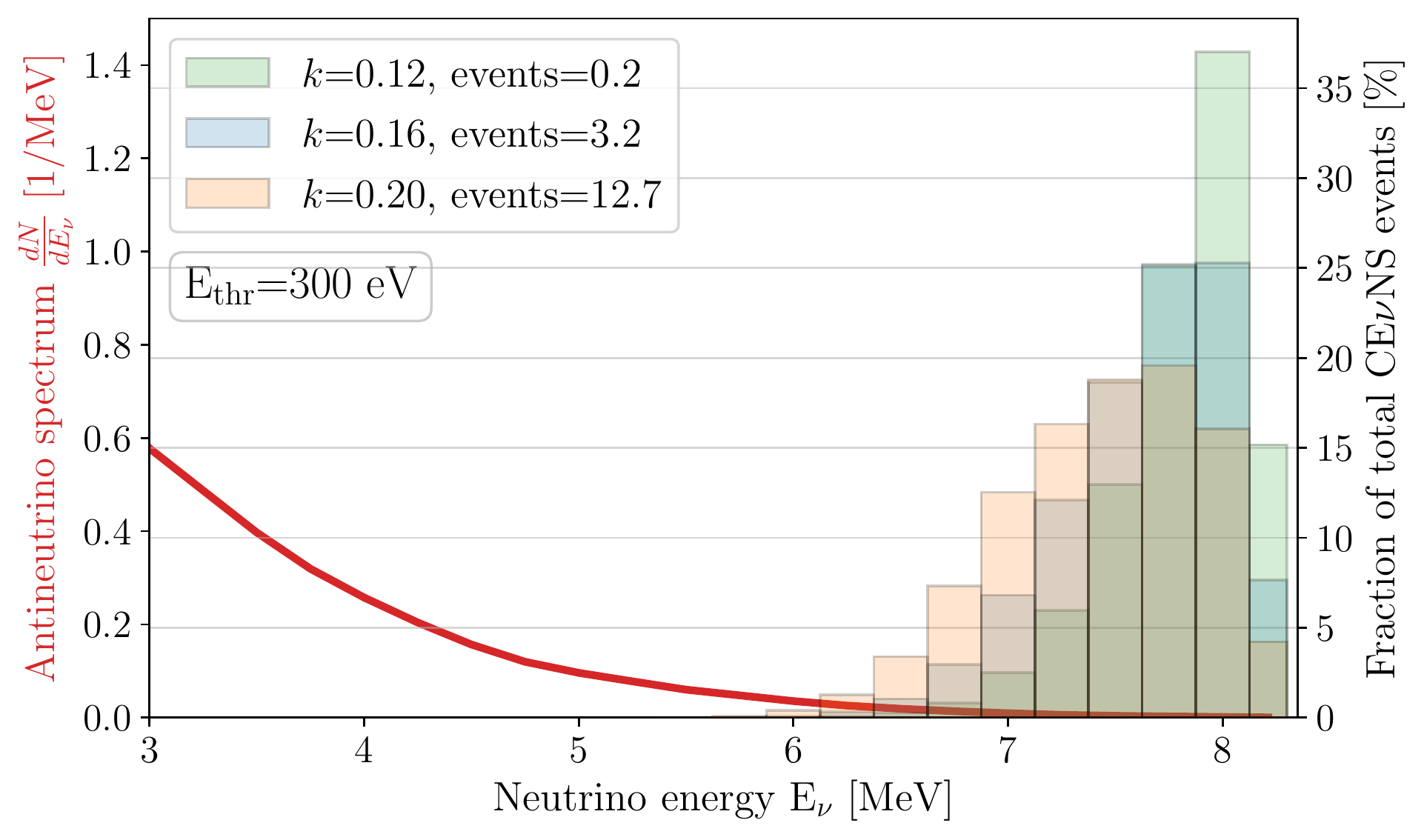}
	\caption[]{Spectral sensitivity of CE$\nu$NS in terms of neutrino energy $E_{\nu}$ for \textsc{Conus} detector C1 and a threshold energy of $E_{\mathrm{thr}}=300$\,eV.
		The underlying reactor antineutrino spectrum is given in red, while the event fraction is indicated for three different $k$ parameters of the modified Lindhard model.
		The figure is taken from Ref.~2.
	}
	\label{fig:spectral_sens}
\end{figure}


In addition to the BSM aspects discussed, CE$\nu$NS will provide further possibilities for investigations within and beyond the SM if the signal is measured with satisfying statistics.
Measurements of the Weinberg angle $\sin^{2}\theta_{W}$ provide further knowledge in low energy regions where uncertainty is still high. 
In particular, any deviation from the generic SM prediction could be a hint towards BSM physics, e.g.\ a light $Z'$-like mediator or a finite neutrino charge radius.
Further, CE$\nu$NS allows a model-independent extraction of the neutron density distribution, while the neutrino sources discussed here contribute knowledge at different momentum transfers.
A direct measurement of a reactor's antineutrino spectrum is another application of this channel. 
Here, CE$\nu$NS exhibits sensitivity to high neutrino energies where uncertainties are still large, cf.~Fig.~\ref{fig:spectral_sens}.

In the context of BSM physics, there are further model classes that can be probed with CE$\nu$NS.
For example, the CE$\nu$NS signal can be used to measure the total neutrino flux and test for oscillation modes into sterile neutrinos.
Although not explicitly designed for this purpose, CE$\nu$NS experiments might contribute further knowledge to the search for eV-mass sterile neutrino discussed in the context of neutrino flux anomalies at short-baselines.
Analogous to the above studies, one can further look for new massive fermions that are created in the CE$\nu$NS process. 
Especially interesting are models that discuss these new fermions in the context of neutrino mass generation or dark matter.
Next to direct interactions, further analyses are possible in terms of newly created particles at $\pi$DAR sources or in nuclear reactor. 
CE$\nu$NS-measuring devices could help to probe portal interactions induced for example by axion-like particles or dark photons. 


\section{Conclusion}

In this talk we argued that CE$\nu$NS opens a new path to high-statistic neutrino physics, where neutrinos can be investigated with ``car''-size detectors.
While this ``new'' interaction channel is going to be probed with the full spectrum of modern detection technologies, reactor and $\pi$DAR experiments represent complementary approaches with their individual strengths and weaknesses.
The recorded and future data allow for various SM and BSM investigations like the measurements of the Weinberg angle or searches for NSIs or light mediating particles. 
The \textsc{Conus} Collaboration has performed several improvements for the recent data collection periods, e.g.\ better environmental stability and a new DAQ system with lower threshold energy and pulse shape capacity.
The analysis of latest data sets from the Brokdorf site is already in preparation and new experimental sites are under discussion.
All in all, CE$\nu$NS allows for an active and bright future in which even flavor-blind neutrino astronomy of the neutrino floor and supernovae seems within reach.
The next generation of CE$\nu$NS experiments in combination with dark matter direct-detection searches will definitely provide a huge playground for phenomenology.


\end{document}